\documentclass[journal]{IEEEtran}

\ifCLASSINFOpdf
\else
\fi

\usepackage{xcolor,colortbl,graphicx}
\definecolor{Gray}{gray}{0.85}
\definecolor{LightCyan}{rgb}{0.88,1,1}
\usepackage{multirow}
\usepackage{booktabs}
\begin{document}
\graphicspath{{./figures/}}
%
\title{Drone technology: interdisciplinary systematic assessment of knowledge gaps and potential solutions}
%
%
%

\author{Evgenii Vinogradov, Sofie Pollin}

\maketitle

\begin{abstract}
Despite being a hot research topic for a decade, drones are still not part of our everyday life. In this article, we analyze the reasons for this state of affairs and look for ways of improving the situation. We rely on the achievements of the so-called Technology Assessment (TA), an interdisciplinary research field aiming at providing knowledge for better-informed and well-reflected decisions concerning new technologies. We demonstrate that the most critical area requiring further  development is safety. Since Unmanned Aerial System Traffic Management (UTM) systems promise to address this problem in a systematic manner, we also indicate relevant solutions for UTM that have to be designed by wireless experts. Moreover, we suggest project implementation guidelines for several drone applications. The guidelines take into account the public acceptance levels estimated in state of the art literature of the correspondent field.



\end{abstract}


\IEEEpeerreviewmaketitle

\section{Introduction}
\IEEEPARstart{C}ivil drones have been a hot topic for a decade. Notably, this interest is shared between industry, academia, business, and civil society. In May 2021, Morgan  Stanley released their long-term prediction \cite{MS} stating that by 2050 (
three decades from now) the total market of urban air mobility (delivery, air taxi, patrolling drones, to name a few) will reach \$9 tn (5-6\% of projected global GDP) or even \$19 tn (11-12\% of projected global GDP) in the "bull case". 

Tech giants like Amazon, Google, and Uber have projects dedicated to drones. Aircraft producers design air taxis. Academia is also very active in this domain: to date, around 300 000 research documents can be found in Scopus as a result of a keyword search string composed of "drone" and its synonyms such as Unmanned aerial vehicle (UAV), Unmanned Aerial System (UAS), Unmanned Aircraft, and Remotely Piloted Aircraft Systems (RPAS). Even when the scope is narrowed down to wireless communication, the interest is significant and well summarized in several tutorial and survey papers \cite{vin_tut,Vin_safe} based on hundreds of selected research articles.

Despite this enthusiasm of different actors, drones are not part of our everyday life. The consensual opinion of industry and academia is that regulation and certification constraints are the main obstacles. For example, even though articles \cite{vin_tut,Vin_safe} are technical, they underline this issue. Furthermore, the authors of \cite{MS} state that \textit{the regulatory requirements for aviation is one of the most underestimated risks}. In fact, in 2021, Morgan Stanley lowered their estimation of the expected market size for 30\% in comparison to their previous Blue paper released in 2018. One of the main reasons was "near-term hurdles related to regulation and certification".

On the other hand, we are far from thinking that the regulatory bodies are malevolent. We realize that managing and regulating technological innovation is a complex task since many actors are involved: from civil society to security agencies, from university researchers to influential and resourceful companies. Partly, the restrictive policies are addressing the public concerns are relatively low trust to the technology. Moreover, unmanned aviation is a new area with many uncertainties and knowledge gaps. 

Since academic researchers are often public servants, we should aim at closing the relevant gaps and address public concerns. By doing this job, we will help policymakers while driving balanced and harmonious technological development. Thus, the first step is to find the relevant problems and then appropriately solve them. 300 000 published research items say that academia is good at finding problems; the absence of drones around says that the problems are not always relevant. This article aims at taking a wider look at the drone technology and assess the problems that are important for all major actors involved in the technology development.   

This work relies on the achievements of the Technology Assessment (TA). TA is an interdisciplinary research field aiming at providing knowledge for better-informed and well-reflected decisions concerning new technologies \cite{gw09}. It provides information that could help the actors (policymakers, citizens, engineers, etc.) in developing their strategies and courses of action \cite{TA_toolkit_98}. TA can also be applied to concrete technical products, processes, services, systems \cite{gw09}. It is used by different agencies and commissions of the European Union, the U.S. Government Accountability Office, national research councils. In China, New Zealand, Italy, Canada and many other countries, TA is widely used in sensitive areas such as healthcare.  

In this article, we use TA to identify relevant problems and assess the current solutions aiming to solve those problems.  By this, we will point to relevant research directions for the wireless communications community. Moreover, using TA, we provide tools for selecting appropriate research and development project scope and framework to increase the project efficiency while considering the project nature, technology readiness, and other relevant factors.   
\section{Background: Technology Assessment}
Though TA roots can be traced back to the technology forecasting studies in the 1950s, there is still no consensual and concise definition. It can be explained by the fact that TA was evolving together with society, views on science and its role in the world. For example, recently TA became an important part of two paradigms: Responsible Research and Innovations (RRI) and Sustainable Development. 

The overall TA philosophy is well framed in \cite{CTA}. It sounds like "to reduce the human cost of trial-and-error learning in society's handling of new technologies, and to do so by anticipating potential impacts and feeding these insights back into decision making, and into actors' strategies." Recently TA became an important part of two paradigms: Responsible Research and Innovations (RRI) and Sustainable Development. 

In \cite{TA_toolkit_98}, a TA-inspired adaptation of the product life-cycle was suggested. Any technology evolves from the idea to maturity and, eventually, becomes obsolete. Logically, each 
stage requires appropriate tools aiming at an efficient transition to the next level of evolution. Moreover, depending on the technology's public perception, some tools can be more efficient than others. 
\begin{figure}
    \centering
    \includegraphics[width=1\columnwidth]{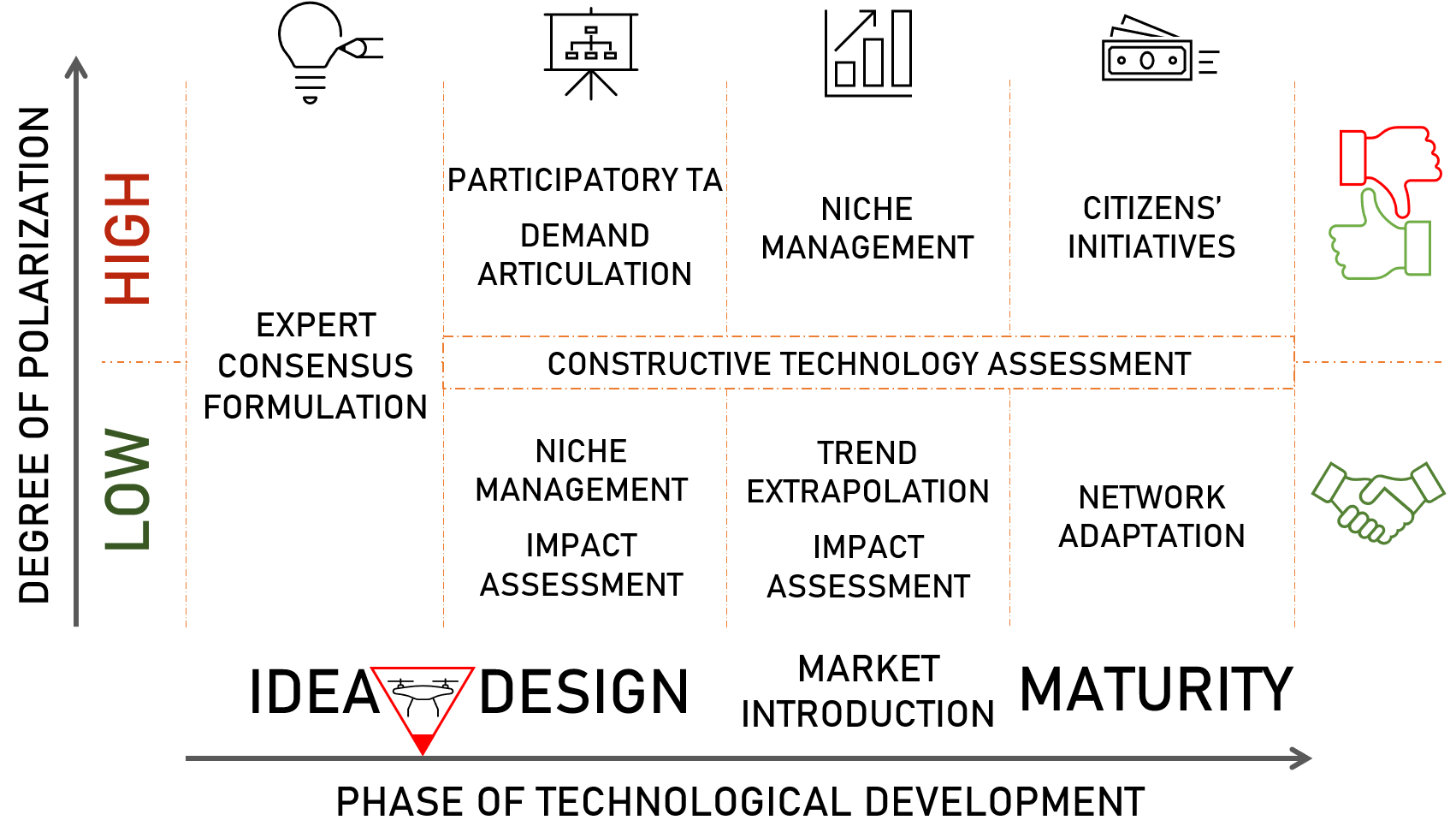}
    \caption{Mapping of TA tools to the phases of technological development. It is advised to apply different tools for technologies causing high and low opinion polarization. Polarizing issues require more participatory approaches.}
    \label{fig:TA_phases}
\end{figure}

Fig.~\ref{fig:TA_phases} shows the mapping of the tools (see \cite{TA_toolkit_98}) on the phases of development and degree of public polarization. Four phases of technological development are defined: 1) idea generation, 2) development/design, 3) market introduction and growth, 4) maturity. Some tools might be inappropriate when opinions in society are highly polarized. In these cases, the preference should be given to the tools with higher inclusion and participation levels since this aims at diminishing contradictions. On the other hand, constructive technology assessment (having a very high level of inclusion) is often seen as less efficient at the late development phases. However, if the polarization level remains high, then the constructive approach is a valid option. 

Note that the current phase of the drone technology development is depicted in Fig.~\ref{fig:TA_phases}. Of course, different drone-applications have different technology readiness levels. Some solutions are becoming commercially available (i.e., phase 3 
Mostly, these products are based on utilizing visual information obtained with a drone-mounted camera (e.g., for construction, infrastructure inspection, law enforcement or entertainment). However, the majority of applications mentioned in \cite{MS,vin_tut} are still rather futuristic. However, while talking about Aerial Base Stations or delivery drones, one may say that the critical mass of research is reached, and the community must start the projects dedicated to 
practical applications.

\section{Technology assessment in identifying relevant problems}
This section is dedicated to knowledge gaps identification and analysis. 
\subsection{Methods of eliciting the experts consensus} 
For a successful transfer to the design phase, the expert consensus is critical. The degree of agreement can be assessed through systematic literature reviews, interviews, conferences/debates and public discussions. 

\textbf{A systematic literature review} (SLR) identifies, selects and critically appraises research in order to answer a clearly formulated question \cite{SLR}. SLR is a reproducible, comprehensive, transparent search conducted over multiple databases and grey literature. The review identifies the type of information searched, critiqued and reported within known timeframes. The search terms, search strategies (including database names, platforms, dates) and limits are included in the review.

\textbf{Delphi method} aims at creating consensus between experts on future developments. The method includes interviewing and anonymously exchanging answers between experts. Anonymity is necessary to estimate future developments without any interference of the social relations between the involved experts. The results of a Delphi are used to explore options for future developments, although materialization 
still depends on specific actions of actors involved. If necessary, the most probable scenario can be defined. \textit{Remark:} Unfortunately, any form of interviewing experts (including Delphi) generally produces biased results. Experts tend to be too optimistic about technological possibilities in the short term (5 to 10 years). However, in the longer term (20 to 50 years), they tend to be too pessimistic.

\textbf{Cross-impact analyses} are used as a variation of Delphi in cases where an event's chance is conditional to other events. The analysis is performed in several steps: i) experts fill in matrices on the chances that an event will occur, given that the other event will (not) occur; ii) the matrices can be manipulated mathematically to calculate the likeliness of series of events. Note that different mathematical tools may be needed if it is expected that the matrices are time-variant.

\subsection{Drone Technology Assessment Results}
In this work, we rely on a TA-inspired systematic literature review \cite{TA_review} and a report \cite{TA_uavs_18} providing an overview of interviews and literature studies. 

Kellerman et al. in \cite{TA_review} reviewed 111 multidisciplinary papers with 2581 relevant quotations. These quotations were subdivided into anticipated barriers (426), potential problems (1037), proposed solutions (737) and expected benefits (381). Here we focus only on problems and solutions. 

\textbf{Problems} ranked by the perceived importance: legal (23.9\%, 248 references); ethical (22.7\%, 235, including threats to privacy 118/235); threats to physical safety (22.0\%, 228); Societal issues (12.8\%, 133); environmental interrelations (7.5\%, 78); economic problems (6\%, 62). In our experience, technical articles mostly express concerns about prohibitive regulations and sometimes safety and security (and privacy). The other issues are much less popular among engineers. 

\textbf{Proposed solutions} have mostly legal and technical nature (27.6\%, 206 and 27\%, 199). Other solutions are targeting public acceptance (14\%, 103), planning and infrastructure (8.1\%, 60), economic factors (6.1\%, 45), safety and security (4.7\%, 35), and environmental aspects (3\%, 22). Note a great 
imbalance between the perceived importance of a problem and the solutions targeting these problems: the most indicative example is that only 4.7\% of the proposed solutions 
target safety that is highlighted as a problem by 22\% of the papers. Privacy solutions are also underinvestigated.
\begin{figure}[]
    \centering
    \includegraphics[width=0.9\columnwidth]{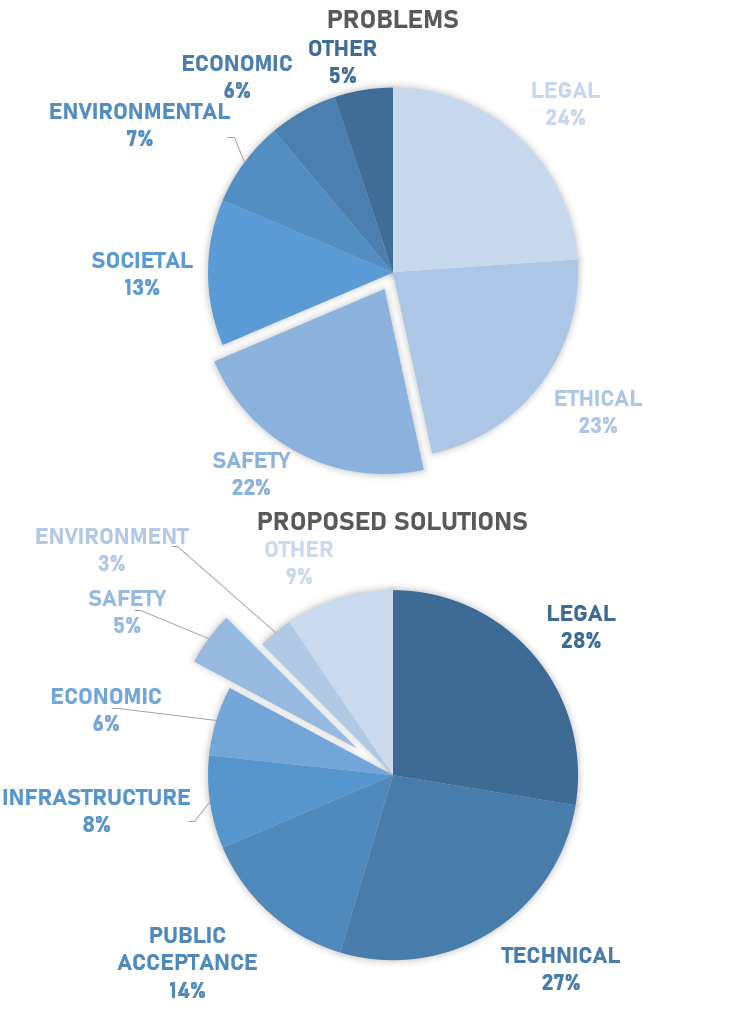}
    \caption{Overview of perceived problems importance and share of proposed solutions. Safety is one of the most important problems; however, it is not reflected in the number of proposed solutions.}
    \label{fig:Prob_sol}
\end{figure}

\begin{figure}[b]
    \centering
    \includegraphics[width=1\columnwidth]{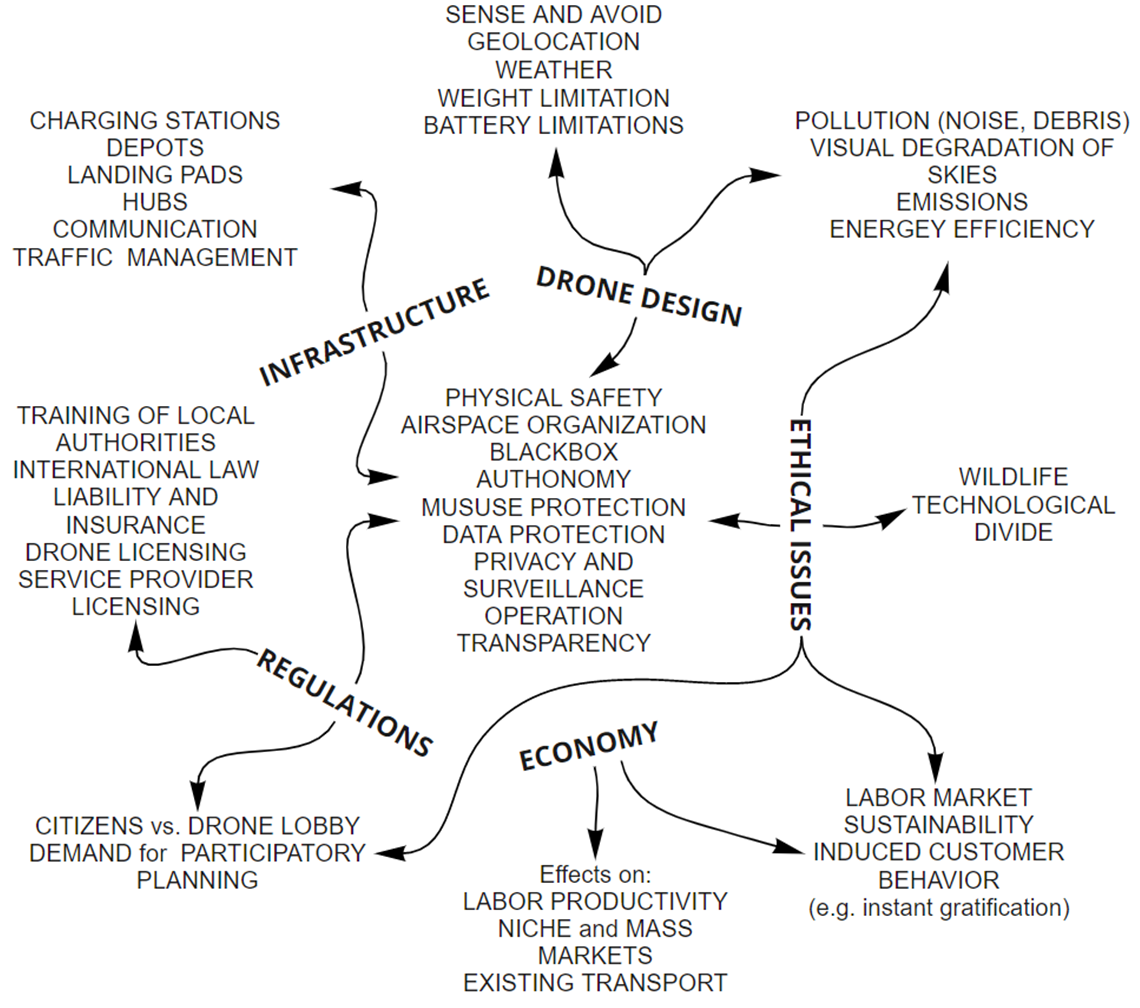}
    \caption{Mindmap of factors influencing the drone technology. Harmonious technology development requires multidisciplinary teams.}
    \label{fig:TA_factors}
\end{figure}

\textbf{A Summary} of the important factors is given in Fig.~\ref{fig:TA_factors}. To create this overview figure, we combined the factors found in \cite{TA_review} with important issues reported in another massive TA study dedicated to UAVs \cite{TA_uavs_18}. It is obvious from the figure that drone technology development is a complex and multidimensional task
requiring efforts from experts from different fields (e.g., from  natural, social sciences, from humanities and regulators). We would like to underline that according to the authors of \cite{TA_review}, the number of proposed solutions often does not correlate with the perceived importance of the problems. 
We identify safety as the least investigated issue.


\subsection{Relevant research directions}
\begin{figure}
    \centering
    \includegraphics[width=0.9\columnwidth]{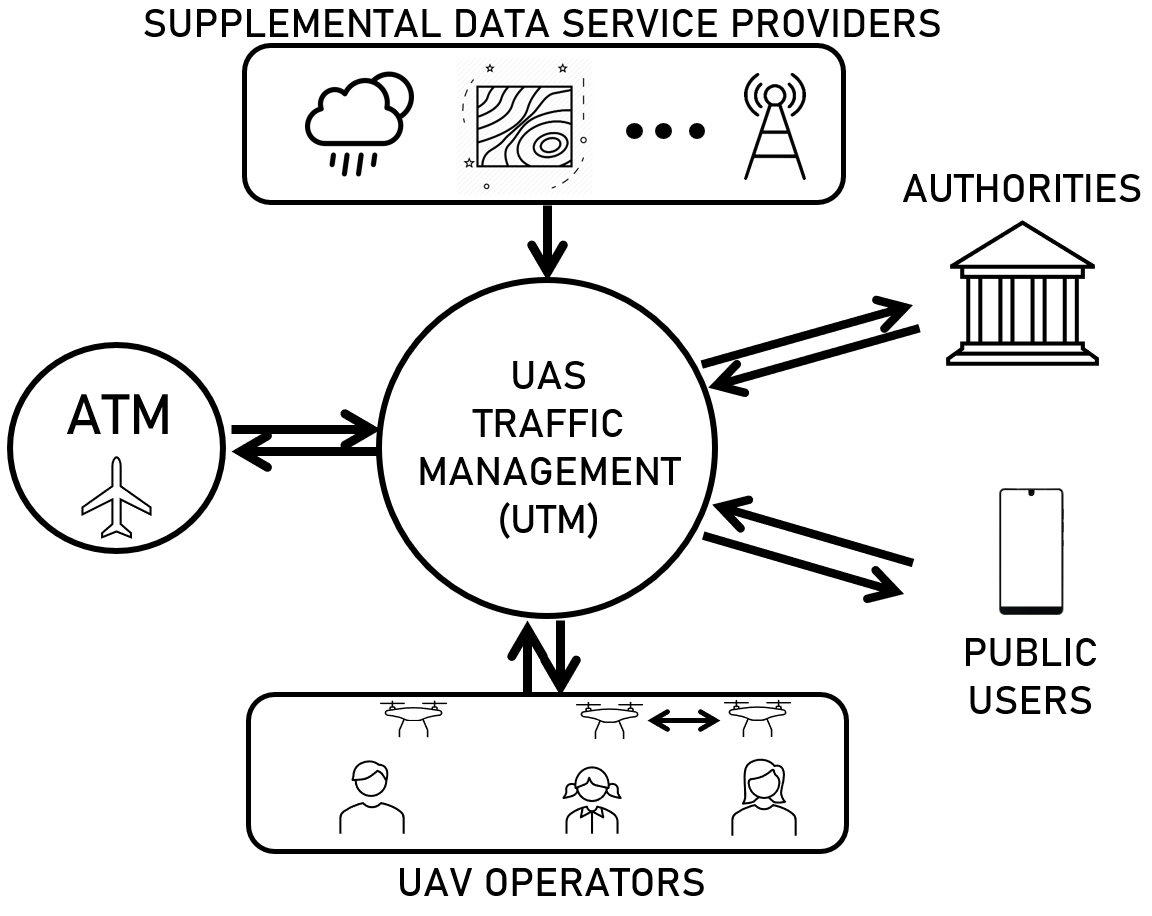}
    \caption{A scheme of UAS Traffic Management system: infrastructure dealing with the majority of factors listed in Fig.~\ref{fig:TA_factors} }
    \label{fig:utm}
\end{figure}

\textbf{UAS Traffic Management} (UTM) systems (see Fig.~\ref{fig:utm} and \cite{Vin_safe}) aim at dealing with the majority of factors listed in Fig.~\ref{fig:TA_factors} through providing a range of services to drone operators (both private and companies), authorities, and other air control bodies such at conventional Air Traffic management (ATM). Moreover, these systems provide a sense of control to the governments, which is vital for loosening the regulations up. Designing these systems (or improving the existing ones) and their components is perhaps the most constructive way technical experts can use to influence the legal constraints.

The main functions of UTM 
are to i) strategically organize the airspace ii) ensure the cooperation of unmanned and manned aviation through a link to ATM iii) ensure transparency of drone operations to citizens iv) control the air traffic through flight permissions v) dynamically adapt drone operations depending on the circumstance (weather conditions, available telecommunication capabilities etc.).

Bauranov and Rakas recently published an overview \cite{UTM_overview} of UTM activities performed by academia, industry, and national and supra-national airspace agencies (e.g., UTM solutions by Nanyang Technological University in Singapore, Amazon and Airbus,  NASA and agencies in Japan, China, Germany as well as the European U-Space concept among many others). Unfortunately, the authors did not include the view of the International Civil Aviation Organization (ICAO) on the recommended common UTM framework with core principles for 
global harmonization presented in \cite{icao_utm}. In \cite{UTM_overview}, we refer the comparison Tables 12 and 13, which compare the proposed solutions based on a subset of factors presented in Fig.~\ref{fig:TA_factors}. Unfortunately, there is no single UTM solution taking into account the full factors list presented in our paper. 
Note that some concepts do not consider UAM (e.g., air taxis) as a part of UTM, but this point of view is less popular.

Though the perfect UTM would account for nearly all factors presented above, in this work,  we focus on the 
conclusion that drone technology lacks safety-oriented solutions.

\textbf{Missing blocks} of UTM systems are indicated by ICAO in \cite{icao_utm}. First of all, this system requires a range of Supplemental Data Services: drone detection/localization/tracking, countermeasures (interception, jamming etc.), micro-weather data providers. Secondly, further research is required to drive the development of the appropriate data standards (e.g. data quality specifications, data protection requirements) and protocols to support UTM safety-related services and the exchange of data. Several standards are needed to accommodate the needs of specific communication flows (UTM-to-ATM, between UTMs, UTM-to-UAV, UTM-to-state authority and others). UTM and ATM systems may have different communications, navigation and surveillance (CNS) requirements for different aircraft types. CNS requirements in UTM may differ from ATM. Data sharing protocols will need to consider State data privacy policies.  

Of course, there are significant cybersecurity risks and 
vulnerabilities that must be taken into consideration. A robust security framework must be established to address potential attacks to communications systems targeting i) C2 Link disruptions, ii) Global Navigation Satellite System (GNSS) jamming or spoofing, iii) manipulation of information exchanged between UAVs and/or UTM systems.

The document \cite{icao_utm} indicated another open problem: UAV separation standards (to avoid collisions and insure optimal airspace capacity) within the UTM system are still missing. Though some work has been previously done \cite{Vin_safe}, these standards should be extended to include the safety margins based on elements such as airspeed, weight and UAV equipment.

\textbf{Cellular technology} is the critical enabler of UTM, as it is pointed out by ICAO in \cite{icao_utm}.  Usage of cellular networks will not be limited by providing reliable controlling and payload links (between UAVs and  operators/UTM). We foresee that many of the missing blocks can be delivered by this technology 
The following cellular-based solutions should be designed/adapted for UTM:
\begin{itemize}
    \item Localization and tracking: vital for verification of the reported coordinates and route. Often UAVs are served by antenna sidelobes which introduce significant localization errors if conventional localization solutions are used.
    \item Detection of non-cooperative (not included in UTM) UAVs: critical service for UTM users awareness. This can be done through passive radio location or simultaneous radar and communication.
    \item Offloading: due to the limited on-board compute power, drones can use the cellular networks to process the visual information (e.g., for visual-based simultaneous localization and mapping).
    \item UAV-to-UAV (U2U) communications are needed for tactical \cite{Vin_safe} deconfliction where drones act autonomously. UAVs may exchange their coordinates, so-called Drone-ID, flight plans (or rather velocity and direction, for security and privacy reasons), type of the vehicle (e.g., to define the deceleration capabilities for an air taxi). This communication can be done through standardized 3GPP sidelink communication (see PC5 interface allowing user-to-user connection without going through the network infrastructure).
\end{itemize}

\section{Technology assessment in shaping the project}
Works \cite{TA_review, TA_uavs_18} suggested that the technologies were almost ready to transit to the second phase. 
Both recommend participatory or constructive technology assessment approaches considering the active involvement of experts, stakeholders and citizens in the research and development projects. This advice is motivated by the opinion that so-called technology-push approach can result in significant backfire from society (we should learn from the 5G conspiracy theories and act differently). This conclusion is confirmed by Morgan Stanley in  \cite{MS} and another reputable consulting company McKinsey in their study performed in 2021 for European Union Aviation Safety Agency (EASA).

In this section, we focus on several tools that can be useful for the second phase of technological development. However, we also pay particular attention to Constructive TA (CTA) since it is a very universal paradigm that can be also used in the following phases. Since "constructive" means a need of continuous communication in this case, we demonstrate the role of communication and identify the relevant parties involved in the communicative process. Finally, we use public acceptance studies to draw our recommendations for selecting appropriate tools from the TA toolkit. Finally, we provide an example of choosing appropriate tools for some drone applications.
\subsection{Overview of TA-inspired project management toolkit}

\textbf{Impact Assessment} analysis is performed in two steps. In-depth analysis is done by the experts in the specific fields that are indicated as being necessary for a particular technology. Next, the technology assessor performs the impact assessment based on expert interviews, brainstorms, and common sense. This process is basically a more advanced version of eliciting expert consensus. Since this tool considers the involvement of a very narrow group of people (or rather a narrow set of possible actors), it is not well suitable for polarizing issues. 

\textbf{Strategic Niche Management} (SNM) is the organization of protected space for a new product or technology where it is easier to experiment with the co-evolution of technologies, user practices, and regulatory structure \cite{Schot2008StrategicNM}. It can be done by setting up a series of experimental settings (niches) in which actors can learn about the design, user needs, cultural and political acceptability \cite{CTA}. Real-life exploitation of a technology (even at a limited scale) allows for the gradual increase of its maturity while minimizing scales of the potential adverse effects (e.g., on general public acceptance).

Note that SNM does not always mean creating a protected market niche since, for many innovations, market niches and user demand are not readily available. This is often the case for the innovations that are not minor variations from the prevailing set of technologies but differ radically. SNM was thus developed for two types of innovation: i) socially desirable innovations serving long-term goals such as sustainability, ii) radical novelties that face a mismatch with regard to existing infrastructure, user practices, regulations, etc. An example of the first type of SNM is the experiments with electric cars in the United States that lead to significant advancement of this technology resulted in the creation of several commercially successful companies such as Tesla. Alternative technologies in the energy sector are a notable example as well. Drone delivery has the potential to become a perfect example of the second type of SNM.

Note that SNM is not a technology-push approach, it instead targets sustainable development requiring interrelated social and technical change. Early works on SNM consider that governments create niches in a top-down fashion. However, Schot and Geels \cite{Schot2008StrategicNM} emphasized the importance of niches emerging through collective enactment (e.g., societal groups). More details on the niche construction, niche-internal processes, policy implications etc. can be found in \cite{Schot2008StrategicNM}. Additionally, the Niche life-cycle (from the creation of a proto-market to privatization) is described in \cite{CTA}

\textbf{Demand Articulation} is a process to make manifest certain latent societal demands for new technology. The most straightforward way to do that is to offer a technology that meets this demand by creating a start-up company or releasing a novel product. Clearly, it is a high-risk option: for example, the first commercially available tablet computer was released in 1989, but these devices were relatively uncommon until the 2010s and the first versions of the iPad.

Another possible strategy for demand articulation is starting up an iterative process between producers, consumers, and knowledge institutions, in which the demand is iteratively better articulated. This process is the first step towards a more participatory approach to technology assessment.

\textbf{Participatory Technology Assessment} (PTA) aims at including interested actors (ranging from experts like researchers or policymakers to lay people) in an innovation process, primarily through discussion meetings, consensus conferences, or workshops. Sometimes other solutions are chosen than expected (by experts) or new, creative solutions are proposed. PTA can use different methods, but the most popular one is Consensus conferences. These are events in which lay people are brought together in a many-day workshop setting to discuss innovation. Involvement of experts is possible if required by the participants. The main goal is to stimulate public debate on a specific subject. The method is particularly important for innovations that involve ethical (including, for example, sustainability) issues resulting in high polarization of opinions. 




\subsection{Constructive Technology Assessment}
Over the last 30 years, Constructive technology assessment (CTA) has been adopted as an approach to technology assessment by many public organizations in the USA, the EU (primarily by Western European and Scandinavian countries such as the Netherlands, Denmark, Norway, Sweden, France, Germany, Austria), several member-states of the Commonwealth of Nations (the UK, Australia, New Zealand), and at the international level (e.g., by Organization for Economic Co-operation and Development - OECD). Since adaptations of the research directions can modify the development course into socially desired directions (e.g., sustainable technology, closing the digital gap etc.), CTA was indicated as one of the main drivers for the Responsible Research and Innovations (RRI) paradigm proposed by the European Commission \cite{Owen13}. Other ideas that are somewhat similar to CTA are socio-technical systems design (STSD), upstream public engagement, value-sensitive design.

CTA (or "user-centric TA") is a development of PTA in which many actors explicitly modulate the process of technology development. CTA is a paradigm in which developers of new technology, consumers, and other relevant actors are brought together in several stages of the development process to discuss research directions, desirable results, future applications, and consumer aspects. In this way, future users' involvement (or even indirectly affected actors) is not limited to expressing their expectations from the technology. It is rather extended to a more democratic process where images of future use of technology are developed and adapted during the process. In other words, the idea is that all involved actors actively \textit{construct} their own knowledge during an active process of learning (through social interactions). In this light, we can say that CTA relies on the theories of knowledge formulated by social constructionists and social constructivists.

For CTA to function correctly, relevant actors should co-produce knowledge and, eventually, innovation. The dynamic nature of the innovation process is central for CTA. Following \cite{CTA}, we can say that this dynamic process happens in several dimensions. In this article, we offer a broader set of dimensions (than in \cite{CTA}) by adapting CTA to a more general paradigm of RRI \cite{Owen13}. The dimensions are:
\begin{itemize}
    \item \textit{Anticipation} involves systematic thinking aimed at increased resilience while revealing new opportunities for innovation and the shaping of agendas for socially-robust risk research. The main questions are what is known, what is likely, what is plausible, what is possible.
    \item \textit{Inclusion} of new voices in research (and in the governance of science in general) aims to build trust within society and increase legitimacy. Particular attention should be paid to the selection of relevant actors as well as intensity, openness, and quality of discussions.
    \item \textit{Reflexivity} is necessary to recognize different roles of actors, but the main goal (for scientists) is to challenge assumptions of scientific amorality and agnosticism. We should not fall back into naive contrast between technology and society. The old-fashioned definition of reflexivity (i.e., based on Poppers's argument that self-referential critique is an organizing principle of science) does not work well in modern society. Emerging techno-skepticism and neo-luddism is the result of this approach to reflexivity.  
    \item \textit{Responsiveness} is the capacity to change development directions, forms, or used tools in response to stakeholder and public values and visions of the future. In other words, the process of innovation design must be modulated. If this does not happen, none of the other dimensions of CTA has any meaning.
\end{itemize}
In practice, the dimensions are non-orthogonal since they may be mutually reinforcing. For example, policymakers or funding agencies' inclusion in research direction discussions can positively affect the responsiveness through development-friendly regulations and goal-optimized access to resources.

We already presented several tools that can be used to develop the dimensions of anticipation (e.g., impact assessment) and responsiveness (e.g., niche management). In the following, we focus on communication and inclusion, aiming at increasing reflexivity.
\subsubsection{Communication in the innovation process}
Traditionally, the role of communication was associated with linear terms such as diffusion and dissemination. In this case, technological development was taken as pre-determined, and it just had to be explained to the public. However, more advanced communication methods are essential in more interactive, constructive, evolutionary or system-oriented modern approaches to innovation design. 

An excellent analysis of the role of communication in innovation processes is given in \cite{Comm_11}. Three main functions of communication and respective communicative strategies (for the complete list refer \cite{Comm_11}) are as follow
\begin{itemize}
    \item Network building 
    \begin{itemize}
        \item Make an inventory of existing initiatives, complemented with stakeholder analysis.
        \item Arrange contacts between disconnected networks with compatible interests.
        \item Manage networks interdependencies.
    \end{itemize}
    \item Support social learning
    \begin{itemize}
        \item Explore and exchange stakeholder perspectives (visualizing the interdependencies).
        \item Elicit uncertainties that hinder development and design collaborative investigation and experimentation to develop common starting points.
        \item Organize regular reflection on process dynamics and satisfaction with outcomes.
    \end{itemize}
    \item Dynamic conflict management
    \begin{itemize}
        \item Identify and propose process facilitators who are credible and trusted by the stakeholders involved.
        \item Steer collaborative research activities to questions relevant to less resourceful stakeholders.
        \item Make stakeholders talk in terms of proposals and counter-proposals
        \item Ensure regular communication with constituents to take them along in the process.
    \end{itemize}
\end{itemize}

Summarizing, current ways of thinking about innovation and communication imply that communication professionals can engage in multiple tasks in the sphere of process preparation, intermediation (agent matching) 
and facilitation. 

As it was demonstrated by the widespread of 5G-related conspiracy theories, media is a powerful tool for changing public opinion about the technology. Consequently, an appropriate choice of media channels and communicative methods is crucial. For example, the linear model mass-media are useful, especially for increasing awareness about the (scientific) consensus. Social media are a powerful tool for fostering greater resonance of new discourses and conversations through 'spreading stories'. Concluding, communication professionals are an essential part of innovation processes.

\subsubsection{Actors}
Schot and Rip \cite{CTA} identified three types of actors. \textit{Technology actors} are those who carry the technological development and invest in it. Examples of these actors are research centers, laboratories, firms, and governmental and commercial funding agencies or technology programs. \textit{Societal actors} are those who anticipate and try to feedback into technological development through regulation, campaigning, educating, etc. Examples of these actors are various societal groups, governmental agencies, regulatory bodies, but the technology actors can combine the two roles and thus short-circuit the feedback.  Finally, \textit{meta-level actors} should facilitate and modulate the interactions between the actors. In some cases, government adjudicating among actors can be seen as an example, but non-governmental communication professionals have equal or higher chances of successfully performing this function \cite{Comm_11}. Moreover, communication professionals have the necessary tools to ensure that the groups of societal actors are representative.

\subsection{Recommendations for selecting right approaches to plan a drone-related project}

\begin{figure}[]
    \centering
    \includegraphics[width=1\columnwidth]{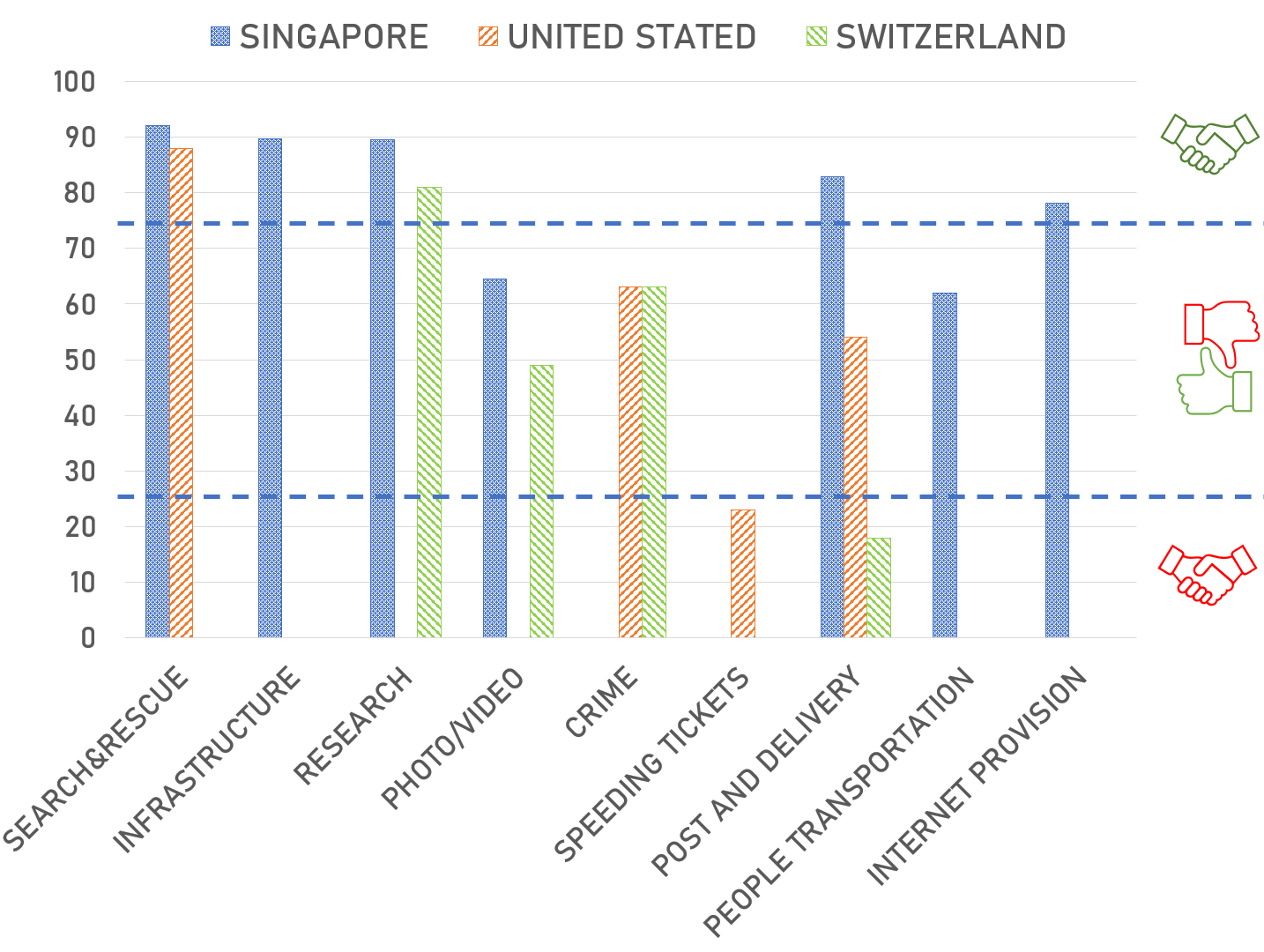}
    \caption{Public acceptance of different UAV applications. The dashed lines indicate 25\% and 75\% used as borders between high and low polarizing issues.}
    \label{fig:PA}
\end{figure}

We suggest using levels of public acceptance as the metric characterizing the degree of polarization. In order to do this, we rely on research articles dedicated to public acceptance of drones. 
Recent article \cite{PA_singapore} presents new results for the public acceptance of the drone technology among Singapore nationals and compares them with similar (though less detailed) studies performed in the US and Switzerland (see Fig.~\ref{fig:PA}).

To summarize, we can say that a high degree of polarization levels (i.e., the acceptance is between 25\% - 75\% indicating no consensus) are observed for applications targeting i) photo/video data collection, ii) fight crimes, iii) delivery (only in the US) and iv) people transportation (onle data for Singapore is available. Consequently, it is more rational to use Demand articulation, PTA, and/or CTA. 

Interestingly, interviewees from Singapore and Switzerland expose a lower level of polarization regarding drone delivery though with different outcomes: the consensus is formed in favour and against the technology, respectively. However, a (accepted) niche application is a good option for both countries. For example, it can be drone delivery of medical samples that is positively seen in both countries. However, Swiss companies will have a longer and more challenging way to reach a successful commercial application. 

For applications targeting i) search and rescue, ii) infrastructure inspection, iii) internet provision, companies and research institutes from Singapore can utilize CTA or rely on an Impact assessment that offers less "communication overhead".

Issuing speeding tickets is the most complex issue: the state authorities might be interested in this solution, however, the citizens are completely against it. This actually creates another kind of polarization (cross-agent polarization). We suggest using CTA (including both interested sides as well as engineers and communication experts) in order to solve this disagreement.

\section{Conclusions}
Whether the reader thinks that TA is a good project structure/implementation practice or not, it is worth paying attention to the problems identified by practitioners of this methodology of eliciting perceived technology flaws/gaps and problems. As we demonstrated, the most critical area requiring further development is safety. Since UTM systems promise to address this problem in a systematic manner, we also indicated relevant solutions for UTM that have to be designed by wireless experts. 

By now the reader has already formed her/his opinion on what kind of projects will bring us to the next level of drone technology development. If you decide to make a project following one of the TA frameworks, you have necessary sources of inspiration in the reference list of this paper.
\section*{Acknowledgment}
This research is supported by the Research Foundation Flanders (FWO), project no. S003817N (OmniDrone) .

\ifCLASSOPTIONcaptionsoff
  \newpage
\fi



%
\bibliographystyle{IEEEtran}
\bibliography{references}

%




\begin{IEEEbiographynophoto}{ }
\textbf{Evgenii Vinogradov} obtained his Ph.D. degree in 2017 from UCLouvain, Belgium. In 2017, Evgenii joined KU Leuven where he is working on wireless communications with UAVs. He actively participated in SESAR (Single European Sky ATM Research) project Percevite.

\textbf{Sofie Pollin} 
is professor at KU Leuven focusing on wireless communication systems. Before that, she worked at imec and UC Berkeley. She has been working on UAV communication since 2013 and is in this context officer of the IEEE Comsoc Aerial Communications ETI.
\end{IEEEbiographynophoto}



\end{document}